# Fundamental Differences Between Application of Basic Principles of Quantum Mechanics on Atomic and Higher Levels[1]

Alexey Nikulov

*Institute of Microelectronics Technology and High Purity Materials, Russian Academy of Sciences, 142432 Chernogolovka, Moscow District, RUSSIA*

**Abstract.** Recent experimental result has revealed that the Bohr's quantization violates symmetry between opposite directions on the macroscopic level. This breach of symmetry is connected with challenge to the law of momentum conservation and the second law of thermodynamics.

## INTRODUCTION

One of the fundamental problem of today is a validity of a direct extrapolation of quantum principles developed for atomic level to the macroscopic level. The most obvious doubt is connected with the contradiction between quantum mechanics and macroscopic realism [1,2]. An assumption on quantum superposition of macroscopically distinct states raises the question: "Is the flux there when nobody looks?" [1] or even "Is the moon there when nobody looks?" [3]. No direct experimental test excluding the hypothesis of macroscopic realism and giving direct evidence of a macroscopic superposition was obtained for the present. In contrast to this there is no doubt about experimental evidence of the Bohr's quantization on the macroscopic level. But an experimental result obtained recently reveals that even this principle can not be extrapolated directly on the macroscopic level without contradiction with some fundamental principles of physics.

## BREACH OF SYMMETRY BETWEEN OPPOSITE DIRECTIONS

There was a logical difficulty in the Bohr's model of atom orbits until electron considered as a particle having a velocity since it was impossible to answer on the question: "What direction has this velocity?" The uncertainty relation and the wave quantum mechanics have overcame this difficulty, removing the velocity direction.

---

[1] The paper is published in the AIP Conference Proceedings Vol. 905, p. 117 (2007) FRONTIERS OF FUNDAMENTAL PHYSICS: Eighth International Symposium FFP8, Madrid, October 17 - 19, 2006

Albert Einstein considered it as weakness of the quantum theory: *The weakness of the theory lies ... in the fact, that it leaves time and direction of the elementary process to "chance"* (the citation from [4]). But thanks of this weakness the Bohr's quantization does not violate symmetry between opposite direction on the atomic level. In contrast to this there is an unambiguous experimental evidence [5] of breach of symmetry on the macroscopic level.

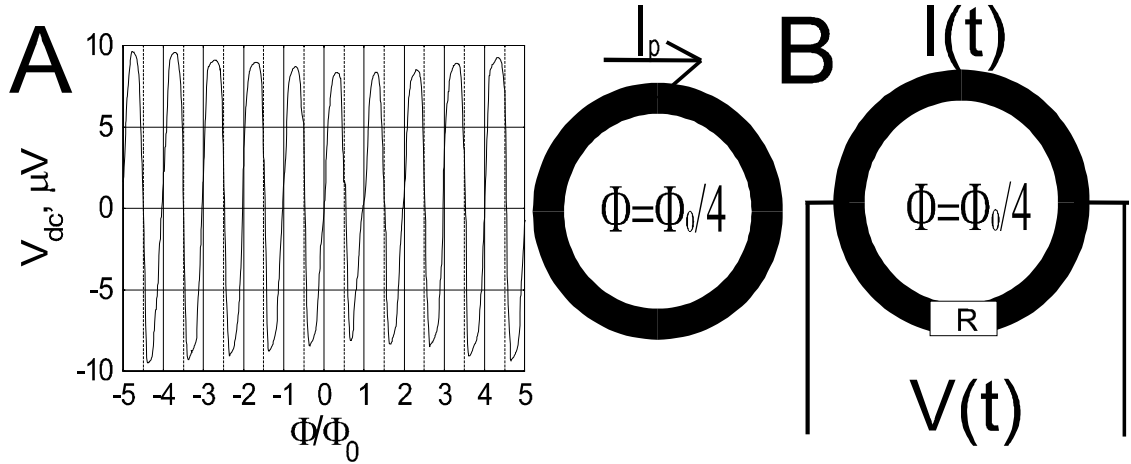

**FIGURE 1.** A) The quantum oscillation of the dc voltage $V_{dc}(\Phi/\Phi_0)$ observed on the semi-ring of asymmetric aluminum ring at $T = 0.97T_c$ in [5]. B) Superconducting states in a ring with closed (at left) and unclosed (at right) wave function.

The periodical change in magnetic field B of sign and value of the dc potential difference $V_{dc}(B)$, Fig.1, observed on semi-rings of an asymmetric aluminum ring below its superconducting transition $T < T_c$ [5] is unambiguous evidence of the periodical change of the direction of the dc electric field $E = -\nabla V_{dc}$ with the B value. We must ask: "Why so can the direction depend on the value?" Any possible answer should be connected with the Bohr's quantization

$$\oint dl p = \oint dl (mv + 2eA) = m\oint dl v + 2e\Phi = n2\pi\hbar \qquad (1)$$

since the period of oscillations, Fig.1, corresponds to the flux quantum $\Phi_0 = \pi\hbar/e$ inside the ring with square S. The quantum oscillations $V_{dc}(BS/\Phi_0) = V_{dc}(\Phi/\Phi_0)$ are evidence of an intrinsic breach of symmetry between right and left directions [6].

## CHALLENGE TO THE LAW OF MOMENTUM CONSERVATION

The dc voltage $V_{dc}(\Phi/\Phi_0)$ can be induced by switching between quantum states with different connectivity of the wave function which is not possible in atom but is observed in superconducting ring [7]. The persistent current $I_p \propto (n - \Phi/\Phi_0)$ (i.e. equilibrium direct current) is observed in the closed superconducting state [8], Fig.1, because of the Bohr's quantization (1). The potential difference $V(t) = RI(t) = RI_p$

exp(-t/$\tau_{RL}$) damping during a relaxation time $\tau_{RL}$ = L/R should be observed on semi-rings after a transition at t = 0 of a ring segment in the normal state with a resistance R, Fig.1. The dc component of the voltage (equal $V_{dc}$ = <V(t)> = <RI(t)> = <RI$_p$ exp(-t/$\tau_{RL}$)> ≈ L<I$_p$>$\omega_{sw}$ at switching with a frequency $\omega_{sw}$ << 1/$\tau_{RL}$ of the ring with a inductance L) can be observed since the pair velocity takes with overwhelming probability minimum permitted value v = (n2$\pi\hbar$ - 2e$\Phi$)/ml = (2$\pi\hbar$/ml)(n - $\Phi/\Phi_0$) [9] and therefore the average value of the persistent current <I$_p$> = s2en$_s$<v> ∝ <n> - $\Phi/\Phi_0$ ≈ n - $\Phi/\Phi_0$ ≠ 0 at $\Phi$ ≠ n$\Phi_0$. The induced dc voltage $V_{dc}$ ∝ <I$_p$> ∝ <n> - $\Phi/\Phi_0$ is the periodical function $V_{dc}(\Phi/\Phi_0)$ in complete agreement with the experiment, Fig.1.

The dc voltage $V_{dc}(\Phi/\Phi_0)$ should be observed both on the lower semi-ring with switched segment, Fig.1, and on the upper semi-ring remaining all time in superconducting state. The latter is challenge to the law of momentum conservation, connected with the breach of the right-left symmetry, since the force $F_e$ = 2eE of the dc electric field E = -$\nabla V_{dc}$ acting on pairs 2e is not compensated by any force.

## CHALLENGE TO THE SECOND LAW OF THERMODYNAMICS

The switching between superconducting states with different connectivity can be induced by both non-equilibrium and equilibrium noise [7,9]. Therefore the dc power $V_{dc}^2$/R observed in the $V_{dc}(\Phi/\Phi_0)$ quantum phenomenon [5] challenges the second law of thermodynamics [6,10].

## ACKNOWLEDGMENTS

This work has been supported by a grant "Quantum bit on base of micro- and nano-structures with metal conductivity" of the Program of ITCS department of RAS, a grant 04-02-17068 of the RFBR and a grant of the Program "Quantum Nanostructures" of the Presidium of Russian Academy of Sciences.